# Zero Dimensional Field Theory of Tachyon Matter

D. D. Dimitrijevic* and G. S. Djordjevic

*Department of Physics, University of Nis, P.O.Box 224, Nis, Serbia, gorandj@junis.ni.ac.yu*
*\*Institute of Physics, Faculty of Sciences, Nis, Serbia, ddrag@pmf.ni.ac.yu*

**Abstract.** The first issue about the object (now) called tachyons was published almost one century ago. Even though there is no experimental evidence of tachyons there are several reasons why tachyons are still of interest today, in fact interest in tachyons is increasing. Many string theories have tachyons occurring as some of the particles in the theory. In this paper we consider the zero dimensional version of the field theory of tachyon matter proposed by A. Sen. Using perturbation theory and ideas of S. Kar, we demonstrate how this tachyon field theory can be connected with a classical mechanical system, such as a massive particle moving in a constant field with quadratic friction. The corresponding Feynman path integral form is proposed using a perturbative method. A few promising lines for further applications and investigations are noted.

**Keywords:** Tachyon, Feynman path integral, String Theory.

**PACS:** 03.65-w, 10., 11.10-z.

## INTRODUCTION

In string theory, when physicists calculate mass of the particles, in some cases, their $mass^2$ turned out to be negative. Such particles are called tachyons. For such a theory vacuum state is generally unstable. A. Sen proposed a field theory of tachyon matter few years ago [1,2]. The action is given as:

$$S = -\int d^{n+1}x V(T)\sqrt{1+\eta^{ij}\partial_i T \partial_j T} ,\quad (1)$$

where $\eta_{00} = -1$, $\eta_{\mu\nu} = \delta_{\mu\nu}$, $\mu,\nu = 1,...,n$, $T(x)$ is the scalar tachyon field and $V(T)$ is the tachyon potential, which unusually appears in the action as a multiplicative factor and has (from string field theory arguments) exponential dependence with respect to the tachyon field:

$$V(T) = e^{-\alpha T(x)} . \quad (2)$$

It is very useful, at least from the pedagogical reason, to understand and to investigate lower dimensional analogs of this tachyon field theory [3].

## ZERO DIMENSIONAL CASE

The corresponding zero dimensional analogue of a tachyon field can be obtained by the correspondence [3]: $x^i \to t$, $T \to x$, $V(T) \to V(x)$. The action reads:

$$S = -\int dt V(x)\sqrt{1-\dot{x}^2} , \quad (3)$$

Corresponding equation of motion, including $V(x) = e^{-\alpha x}$, is:

$$\ddot{x} + \alpha \dot{x}^2 = \alpha , \quad (4)$$

and coincide with the equation for the system under gravity in the presence of quadratic damping:

$$m\ddot{y} + \beta \dot{y}^2 = mg . \quad (5)$$

This equation can be derived from the action:

$$S = -\int dt\, e^{-\frac{\beta y}{m}} \sqrt{1 - \frac{\beta}{mg}\dot{y}^2} . \quad (6)$$

The solution can be found perturbatively:

$$y(t) = y_0(t) + y_1(t), \qquad (7)$$

where $y_0(t)$ is solution of Eq. (5) for $\beta = 0$, and $y_1(t)$ is obtained from the same equation after inserting $y_0(t)$ and neglecting all non linear terms:

$$\ddot{y}_1 + at\dot{y}_1 = -bt^2, \qquad (8)$$

where $a = \frac{2\beta g}{m}$, $b = \frac{\beta g^2}{m}$. For $y_0(0) = 0$ and $\dot{y}_0(0) = 0$ the final solution is given by:

$$y(t) = \frac{g}{2}t^2 + \frac{b}{2a}t^2({}_2F_2[1,1;\frac{3}{2},2;-\frac{at^2}{2}]-1), \quad (9)$$

where ${}_2F_2[1,1;\frac{3}{2},2;-\frac{at^2}{2}]$ is hypergeometric function. For small $t$ it gets quite simple form:

$$y(t) = \frac{g}{2}t^2 - \frac{\beta g^2}{12m}t^4. \qquad (10)$$

This solution with nontrivial boundary conditions can be useful to get simpler-quadratic action for tachyons. Details will be presented elsewhere.

## FEYNMAN PATH INTEGRAL

According to Feynman's idea [4], dynamical evolution of the system is completely described by the kernel $K(y'',T;y',0)$ of the unitary evolution operator $U(0,T)$, where $y'', y'$ are initial and final positions and $T$ is ``total`` time:

$$K(y'',T;y',0) = \int Dy e^{\frac{2\pi i}{h}\int_0^T Ldt}. \qquad (11)$$

There is very useful semi-classical expression for the kernel if the classical action $\bar{S}(y'',T;y',0)$ is polynomial quadratic in $y'$ and $y''$ (which holds for both real and p-adic number fields [5]):

$$K(y'',T;y',0) = \left(\frac{i}{h}\frac{\partial^2 \bar{S}}{\partial y' \partial y''}\right)^{1/2} e^{\frac{2\pi i}{h}\bar{S}(y'',T;y',0)}. \qquad (12)$$

One can go back to Eq. (6) and for very small $\beta$, it leads to the new form of action (6):

$$S \underset{\beta \to 0}{\to} S' = -\int dt[\frac{\beta}{2mg}\dot{y}^2 + \frac{\beta}{m}y - 1], \quad (13)$$

This action is quadratic with respect to velocity, and standard procedure can be engaged for the path integral.

## CONCLUSION

Sen`s proposal [1,2] and similar conjectures (see, e.g., [6]) have attracted important interests among physicists. Our understanding of tachyon matter, especially its quantum aspects is still quite pure. Perturbative solutions for classical particles analogous to the tachyons offer many possibilities for further investigations and toy models in quantum mechanics, quantum and string field theory and cosmology on archimedean and nonarchimedean spaces [5].

## ACKNOWLEDGMENTS

The research of both authors is supported by the Serbian Ministry of Science and Technology Projects No. 144014 and No. 141016. The financial support of the UNESCO-ROSTE under the Project ``Southeastern European Network in Mathematical and Theoretical Physics`` (SEENET-MTP) No. 8759145 is also kindly acknowledged. We would like to thank S. Kar, J. Jeknic and Lj. Nesic for many fruitful discussions.